\documentclass[a4paper,11pt,noeprint]{article}
\usepackage{pos}

\usepackage{graphicx}
\usepackage{amsmath}
\usepackage{epstopdf}

\title{Contribution title}
\title{Studying the nature of the lightest charmed axial mesons via femtoscopy}

\author*[b]{Luciano M. Abreu}
\author[a]{K.~P.~Khemchandani}
\author[c]{A.~Mart\'inez~Torres}
\author[c]{F.~S.~Navarra}

\affiliation[a]{Universidade Federal de S\~ao Paulo, \\
C.P. 01302-907, S\~ao Paulo, Brazil}

\affiliation[b]{Instituto de F\'isica, Universidade Federal da Bahia, \\
40210-340, Salvador, BA, Brazil}

\affiliation[b]{Instituto de F\'isica, Universidade de S\~ao Paulo, \\
 C.P. 05389-970, S\~ao Paulo,  Brazil}

\emailAdd{luciano.abreu@ufba.br}
\emailAdd{kanchan.khemchandani@unifesp.br}
\emailAdd{amartine@if.usp.br}
\emailAdd{navarra@if.usp.br}

\abstract{
In this work we discuss how femtoscopic analysis can shed light on the nature of the two lightest axial charmed mesons, denominated as  $D_1(2430)$ and $D_1(2420)$, whose masses are similar but widths are different. Their properties are reasonably described taking into account meson-meson  coupled channel dynamics and a bare quark-model pole constituting the lowest-order amplitudes. Two different bare quark-model states are used in order to accommodate the different scattering lengths coming from the lattice QCD calculations for the $D\pi$ and $D^*\pi$ systems and the data from the ALICE Collaboration on the $D\pi$ system. The amplitudes are employed as inputs to determine the correlation functions for the $D^{*}\pi$ and $D\rho$ channels and identify the signatures associated with the lowest-lying axial charmed mesons. 
}

\FullConference{10th International Conference on Quarks and Nuclear Physics (QNP2024)\\
8-12 July, 2024\\
Barcelona, Spain\\}

\begin{document}
\maketitle

\section{Introduction}

Fits made to the experimental data on the $D^*\pi$ invariant mass distribution of the decay $B^- \to D^{*+} \pi^- \pi^-$ support the experimental evidence of the two lightest, charm, axial mesons~\cite{LHCb:2015tsv,Belle:2003nsh}. Their average values of mass and width, as given by the Particle Data Group (PDG)~\cite{PDG2022}, are:
\begin{align}\nonumber
D_1(2420):&~~~ M=2422.1\pm 0.6 \text{ MeV,    } \Gamma=31.3\pm1.9\text{ MeV.}\\
D_1(2430):&~~~ M=2412\pm 9 \text{ MeV,    } \Gamma=314\pm29\text{ MeV.}\label{pdgvalues}
\end{align}

On theoretical grounds, several attempts have been made to simultaneously describe the properties of the $D_1$ states, but without a consensus on the topic. Different approaches based on distinct interpretations of these states have been employed, generating different results; as examples we refer the reader to the Refs.~\cite{Godfrey:1985xj,DiPierro:2001dwf,Bardeen:2003kt,Colangelo:2004vu,Mehen:2005hc,Ni:2021pce,Kolomeitsev:2003ac,Guo:2006rp,Gamermann:2007fi,Coito:2011qn,Burns:2014zfa,Ferretti:2015rsa,Du:2017zvv,Abreu:2019adi,Malabarba:2022pdo,Torres-Rincon:2023qll,Khemchandani:2023xup}. Thus,  alternative approaches  proposing tools and observables that in principle might distinguish the nature of the $D_1$ states are needed. In this scenario, we investigate if femtoscopic studies can play such a role. 

To do that, we perform an extension of the framework introduced in Refs.~\cite{Gamermann:2007fi,Malabarba:2022pdo}. We notice that a narrow pole is found in the previous work~\cite{Gamermann:2007fi} from meson-meson coupled channel dynamics, which couples mainly to the $ D\rho$ channel. Improvements have been implemented in the subsequent work~\cite{Malabarba:2022pdo}, with the inclusion of additional interaction diagrams to the lowest order amplitude, yielding a mass and a width in agreement with those of $D_1(2420)$, but not for the case of the broader resonance $D_1(2430)$. In view of this, we consider the addition of a quark-model pole to the lowest order $D^*\pi$ amplitude of Ref.~\cite{Malabarba:2022pdo} as an attempt to find a better description of the mass and width of the $D_1(2430)$.    
To test the reliability of this approach we constrain the amplitudes to reproduce the different scattering lengths ($a^{(1/2)}$) estimated from the lattice QCD calculations for the $D\pi$ and $D^*\pi$ systems~\cite{Mohler:2012na} and the data from the ALICE Collaboration on the $D\pi$~\cite{ALICE:2024bhk} via distinct choices of the bare quark model state.
%(see also Refs.~\cite{Alice2,Alice3,Alice4}). 
In this last case the value determined for the $a^{(1/2)}$ is different to that of previous theoretical studies~\cite{Liu:2012zya,Guo:2018tjx,Guo:2009ct,Abreu:2011ic,Geng:2010vw}. Here we invoke  heavy-quark symmetry arguments and assume that the  $D^*\pi$ and  $D\pi$ systems have similar $a^{(1/2)}$. Thus, we contemplate both possibilities of the $a_{D^*\pi}^{(1/2)}$. After that, the amplitudes are employed in the correlation functions (CFs) for the $D^{*}\pi$ and $D\rho$ channels to identify the signatures associated with the lightest $D_1$ states. 
 
 \section{Scattering amplitudes}

We start by introducing the scattering amplitudes describing the interactions between vector and pseudoscalar mesons, which are  obtained from a Lagrangian based on a broken $SU(4)$ symmetry~\cite{Gamermann:2007fi,Malabarba:2022pdo,Khemchandani:2023xup}.  Specifically, the lowest-order kernels are given by
 \begin{align}
 V_{ij}=\frac{C_{ij}}{4f^2}\left(s-u\right) \vec \epsilon\cdot\vec \epsilon^\prime,\label{tamp}
 \end{align}
where $f$ is the pion decay constant, taken to be 93 MeV, $s$, $u$ are Mandelstam variables, $\epsilon \left(\epsilon^\prime\right)$ represents the polarization vector for the incoming (outgoing) vector meson, and $C_{ij}$ are constants for different $i,~j$ initial, final states, given in Table I of Ref.~\cite{Khemchandani:2023xup} for the isospin 1/2 configuration. Besides Eq.~(\ref{tamp}), we also consider contributions coming from a pseudoscalar exchange through box diagrams  as obtained in Ref.~\cite{Malabarba:2022pdo} for the $D\rho \to D^*\pi \to D\rho$. 

As discussed in Ref.~\cite{Khemchandani:2023xup}, 
two poles are found from the dynamics, one close to real axis which agrees with the properties of $D_1(2420)$ and another whose mass and width are close but not in agreement with those of $D_1(2430)$.  To better describe the properties of $D_1(2420)$ and $D_1(2430)$ simultaneously, we add a bare quark-model pole to the lowest order amplitude for the $D^*\pi$ channel 
\begin{align}
V_{QM}=\pm\frac{g^2_{QM}}{s-M_{QM}^2},\label{vqm}
\end{align}
where the mass $M_{QM}$ can be taken from different  quark model calculations~\cite{Godfrey:1985xj,Ferretti:2015rsa,Abreu:2019adi} and $g_{QM}$ is fitted to obtain a fair agreement between the wider pole and the properties of $D_1(2430)$. 

The scattering amplitudes are obtained by solving the Bethe-Salpeter equation 
\begin{align}
T=V+VGT,\label{BSE}
\end{align}
where we recall that the kernel $V$ is given in Eq.~(\ref{tamp}) (plus the bare quark-model pole in Eq.~(\ref{vqm}) for the $D^*\pi$ channel); and $G$ is the meson-meson loop function, which is diagonal in the channel space, and within the dimensional regularization scheme has the parameters $\mu=1500, a=-1.45$~\cite{Khemchandani:2023xup}.

In Table~\ref{tableAB} we show the poles found with the solutions of Eq.~(\ref{BSE}), the values of the scattering length and the coupling  for the most relevant channels $D^*\pi$, $D\rho$, $\bar K D^*_s$ and $D_s \bar K^*$. The findings are presented for two different bare quark-model states $V_{QM}$ to accommodate the different isospin 1/2 scattering lengths coming from the lattice QCD calculations~\cite{Mohler:2012na} (Model A) and the data from the ALICE Collaboration~\cite{ALICE:2024bhk} (Model B) as previously mentioned. 
We can see that in both scenarios the estimated values of the $D^*\pi$ scattering length are in agreement with the mutually conflicting ones reported in Refs.~\cite{Mohler:2012na} and~\cite{ALICE:2024bhk}. 
In the case of Model A, the $D^*\pi$ amplitude shows a peak on the real axis around 2304 MeV, with a width of around 160 MeV. Such a width is more in agreement with the lower limit determined by the Babar Collaboration~\cite{BaBar:2006ctj}. On the other hand, for the Model B a broad bump is found on the real axis around 2436 MeV with a full width at half maximum being $\sim$ 311 MeV, which is closer to the findings of the LHCb and Belle Collaborations~\cite{LHCb:2015tsv,Belle:2003nsh} [as also given in Eq.~(\ref{pdgvalues})].
We now study how such features show up in the CFs.

\begin{table}[ht!]
\centering 
\caption{Values of the poles, isospin 1/2 scattering lengths and the couplings (represented as $g$) of the two states $D_1$ for the different channels. $(M,\Gamma)_{D^*\pi}$ represents the peak in the modulus squared amplitude for the $D^*\pi$ channel. The findings are presented for the two choices  of $V_{QM}$ (Models A and B). }\label{tableAB}
\begin{tabular}{cccc}
\textbf{Model A }&  &  & \\ \hline
 $ V_{QM} $ & Pole 1 (MeV) & Pole 2 (MeV) & $(M,\Gamma)_{D^*\pi}$ (MeV) \\
 $ -\frac{(6000\ \text{MeV})^2}{s-(2440\ \text{MeV})^2} $ & $(2428 - i 16 )$  &  $(2268 - i 100 )$  & $(2304, 160)$ \\ \hline
 & $a^{(1/2)}$ \text{(fm)}& $g_{D_1(2430)}$ (MeV) &$g_{D_1(2420)}$ (MeV) \\

$D^*\pi$& $-0.20$ & $7250-i4995$&$-233+i5$ \\
$D\rho$& $0.44-i0.18$&$-521-i355$&$15144+i356$ \\
$\bar K D_s^*$&$0.00-i0.12$&$4534-i3612$&$-247-i177$\\
$D_s\bar K^*$&$0.00-i0.12$&$2-i55$&$-8739+i90$ \\
\hline
\hline
\textbf{Model B }&  &  & \\ \hline
 $ V_{QM} $ & Pole 1 (MeV) & Pole 2 (MeV) & $(M,\Gamma)_{D^*\pi}$ (MeV) \\
 $ -\frac{(10000\ \text{MeV})^2}{s-(2370\ \text{MeV})^2} $ & $(2428 - i 16 )$  &  $(2218 - i 218 )$  & $(2436, 311)$ \\ \hline
 & $a^{(1/2)}$ \text{(fm)}& $g_{D_1(2430)}$ (MeV) &$g_{D_1(2420)}$ (MeV) \\

$D^*\pi$& $0.1$ & $5199-i3577$&$92-i105$\\
$D\rho$& $0.45-i0.18$&$-248-i91$&$14987+i232$\\
$\bar K D_s^*$&$0.00-i0.12$&$3806-i2249$&$186-i39$\\
$D_s\bar K^*$&$0.00-i0.12$&$-34-i36$&$-8668+i151$\\
\end{tabular}
\end{table}

%%%%%%%%%%%%%%%%%%%%%%%%%%%%%%%%%%%%%%%%%%%%%%%%%%%%%%%%%%%%%%%%%%%%
%%%%%%%%%%%%%%%%%%%%%%%%%%%%%%%%%%%%%%%%%%%%%%%%%%%%%%%%%%%%%%%%%%%%
\section{Correlation Functions}\label{sec-corrfunction}
%%%%%%%%%%%%%%%%%%%%%%%%%%%%%%%%%%%%%%%%%%%%%%%%%%%%%%%%%%%%%%%%%%%%
%%%%%%%%%%%%%%%%%%%%%%%%%%%%%%%%%%%%%%%%%%%%%%%%%%%%%%%%%%%%%%%%%%%%

%%%%%%%%%%%%%%%%%%%%%%%%%%%%%%%%%%%%%%%%%%%%%%%%%%%%%%%%%%%%%%%%%%%%
%%%%%%%%%%%%%%%%%%%%%%%%%%%%%%%%%%%%%%%%%%%%%%%%%%%%%%%%%%%%%%%%%%%%
%%%%%%%%%%%%%%%%%%%%%%%%%%%%%%%%%%%%%%%%%%%%%%%%%%%%%%%%%%%%%%%%%%%%
%%%%%%%%%%%%%%%%%%%%%%%%%%%%%%%%%%%%%%%%%%%%%%%%%%%%%%%%%%%%%%%%%%%%

The femtoscopic analysis is based on the estimation of the correlation functions (CFs). We adopt the framework summarized in Refs.~\cite{Vidana:2023olz,Feijoo:2023sfe,Albaladejo:2023pzq,Torres-Rincon:2023qll}, in which the generalized coupled-channel CF for a specific channel $i$ reads
\begin{eqnarray}
C_i(k)  =  1 + 4 \pi \theta (q_{max} - k) \int_{0}^{\infty} d r r^2 S_{12}(\vec{r}) \left( \sum_j w_j \vert j_0(kr) \delta_{ji} + T_{ji}(\sqrt{s}) \widetilde{G}_j(r; s) \vert^2 - j_0^2(kr) \right) ,  
%\nonumber \\
\label{cf2}
\end{eqnarray}
where $ w_j $ is the weight of the observed channel $ j $ (we use $ w_j = 1$); $ j_{\nu}(kr) $ is the spherical Bessel function; $E = \sqrt{s}$ is the CM energy; the relative momentum of the channel is $k = \lambda^{1/2} (s,m_1^2,m_2^2)/(2\sqrt{s})$ ($\lambda$ being the K\"allen function and $m_1, m_2$ the masses of the mesons in the channel $i$); $ T_{ji} $ are the elements of the scattering matrix encoding the meson–meson interactions discussed in the previous section; and the $\widetilde{G}_j(r; s)$ function is defined as 
\begin{eqnarray}
\widetilde{G}_j(r; s) & = & \int_{\vert \vec{q} \vert < q_{max} } \frac{d^3 q}{(2\pi)^3} \frac{\omega_1^{(j)} + \omega_2^{(j)} }{2 \omega_1^{(j)} \omega_2^{(j)} } \frac{j_0(qr)}{s - \left( \omega_1^{(j)} + \omega_2^{(j)} \right)^2 +i \varepsilon} ,   
\label{gtilde}
\end{eqnarray}
with $ \omega_a^{(j)} \equiv \omega_a^{(j)}(k) = \sqrt{k^2 + m_a^2}$ being the energy of the particle $a$, and $q_{max}$ being a sharp momentum cutoff introduced to regularize the $r \to 0$ behavior.  We choose $q_{max} = 700 \  \mathrm{MeV}$. Here we employ a source function parametrized as a static Gaussian normalized to unity, i.e., $S_{12}(\vec{r})   =   \exp{\left[ -r^2/ (4 R^2) \right]} / \left[ \left(  4 \pi \right)^{\frac{3}{2}} R^3  \right], $ where $R$ is the source size parameter. The relations between the CF's in the particle basis and the isospin basis are (analogously for the $D \rho $ system)
\begin{eqnarray}
C_{ D^{*0} \pi^+}  & \equiv  & C_{D^{*0} \pi^+ \to D^{*0} \pi^+} + C_{D^{*+} \pi^0 \to D^{*0} \pi^+}   = \frac{2}{3}  C_{D^{*} \pi}^{\left(\frac{1}{2}\right)} + \frac{1}{3} C_{D^{*} \pi}^{\left(\frac{3}{2}\right)}  \ , \nonumber \\
C_{ D^{*+} \pi^0}  & \equiv  & C_{D^{*0} \pi^+ \to D^{*+} \pi^0} + C_{D^{*+} \pi^0 \to D^{*+} \pi^0} =  \frac{1}{3}  C_{D^{*} \pi}^{\left(\frac{1}{2}\right)} + \frac{2}{3} C_{D^{*} \pi}^{\left(\frac{3}{2}\right)}  \ .
\label{rel2}
\end{eqnarray}
The CFs for the isospin $3/2$ configuration are obtained from the $T^{\left( \frac{3}{2} \right)}$ amplitudes shown in Ref.~\cite{Khemchandani:2023xup}.

\begin{figure}[!htbp]
    \centering
\includegraphics[width=0.42\textwidth]{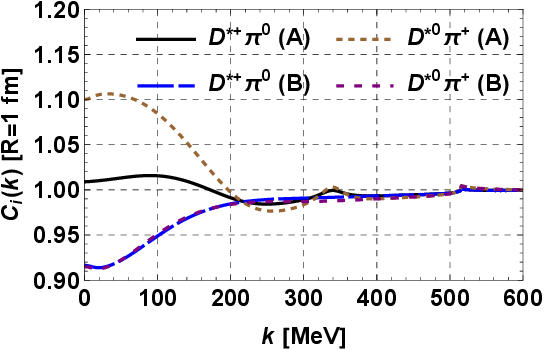}
\includegraphics[width=0.42\textwidth]{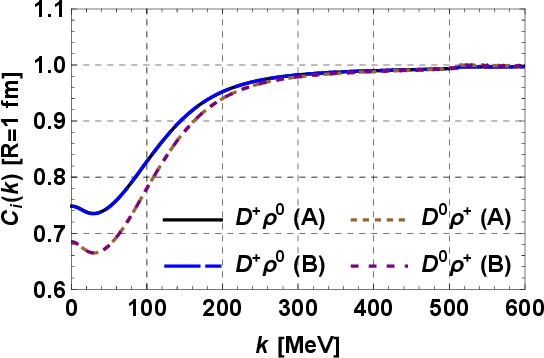}
    \caption{CFs for the physical $D^{*} \pi$ and $ D \rho $ states defined in Eq.~(\ref{rel2}) in both cases of models A and B, taking the source size parameter $R=1$ fm. 
}
    \label{fig:cf12partbasis}
\end{figure}

We show in Fig.~\ref{fig:cf12partbasis} the CFs for the  $D^{*} \pi$ and $ D \rho $ states in the particle basis defined in Eq.~(\ref{rel2}), for both  models A and B. One can notice the distinct behavior of the $C_{D^* \pi}^{(1/2)}(k)$ when models A or B are employed. In the case of model A, the features of the  $T_{D^* \pi,D^* \pi}^{(1/2)}$ amplitude are more notable in the channel $D^{*0} \pi^+$, because of the bigger weight of the $I=1/2$ channel in its wave function. 
At threshold, the attractive character of this channel and the negative scattering length yields $C_{D^* \pi}^{(1/2)}(k=0) > 1$. As $k$ increases, a moderate minimum and a bump are found in the region $220 \lesssim k \lesssim 360 \ \mathrm{MeV}$, reflecting essentially the behavior of the $T_{D^* \pi,D^* \pi}^{(1/2)}$ amplitude, since the other contributions $T_{D^* \pi,D \rho}^{(1/2)}$ and $T_{D^* \pi, \bar K D_s^*}$ are negligible~\cite{Khemchandani:2023xup}. In this sense,   the minimum (bump) at $k \gtrsim 250 \ \mathrm{MeV}$ ($k \simeq 340 \ \mathrm{MeV}$) is associated with the broad peak (dip) in $T_{D^* \pi,D^* \pi}^{(1/2)}$ at $\sqrt{s} \sim 2304 \ \mathrm{MeV}$ ($\sqrt{s} \sim 2405 \ \mathrm{MeV}$). Thus, the CF encodes the manifestation of the interference between the poles present in $T_{D^* \pi,D^* \pi}^{(1/2)}$. Also, a cusp at $k \simeq 518 \ \mathrm{MeV}$ is seen, coming from the effect of the $\bar K D_s^*$ threshold. 
In contrast, for model B there is no sizable difference among the channels $D^{*0} \pi^+ $ and $ D^{*+} \pi^0$, due to the similarity among $C_{D^* \pi}^{(1/2)}(k)$ and $C_{D^* \pi}^{(3/2)}(k)$. 
At the threshold, model B generates $C_{D^* \pi}^{(1/2)}(k=0) \lesssim 1$, which is compatible with the result expected when $a_{D^* \pi}^{(1/2)} = 0.1 \ \mathrm{fm} < 2.3 R$. After that, the CF slightly increases with $k$, and presents a plateau, which comes from the interference between the states, and goes to one. As in the former model, the full CF expresses the behavior of the $T_{D^* \pi,D^* \pi}^{(1/2)}$ amplitude.

Now we move on to the scenario of $ D \rho $, whose scattering length has an imaginary component. We do not see large differences in the $C_{D \rho}^{(1/2)}(k)$ obtained considering the models A and B. 
Since $Re[a_{D \rho}^{(1/2)}] = 0.44 \ \mathrm{fm} < 2.3 R$, then one can expect that  $C_{D^* \pi}^{(1/2)}(k=0) < 1 $. 
However,  when compared with the results for $D^* \pi$ in model B, the CF experiences a more prominent dip. It may be interpreted as the influence of the narrow state present in the $T_{D \rho,D \rho}^{(1/2)}$ amplitude below the $D \rho$ threshold, which provides the relevant contribution.
The difference coming from the isospin weights produces $C_{ D^{+} \rho^0}(k)$ closer to one at threshold than $C_{ D^{0} \rho^+}(k) $.
Hence, one can conclude that the $D^{*0} \pi^+$ and $ D^{0} \rho^+ $ channels are more appropriate to test both models.

To summarize, the main conclusion of this work is that $C_{D^* \pi}(k)$ and $C_{D \rho}(k)$ might encode signatures of the $D_1(2430)$ and $D_1(2420)$ states when smaller sources are considered. Accordingly, this study provides a framework compatible with the existence of both broad and narrow states if the measured genuine CFs present similar behavior to those obtained here. In view of this, new data from high precision experiments would be welcome in order to confront them with our predictions.

%%%%%%%%%%%%%%%%%%%%%%%%%%%%%%%%%%%%%%%%%%%%%%%%%%%%%%%%%%%%%%%%%%%%
%%%%%%%%%%%%%%%%%%%%%%%%%%%%%%%%%%%%%%%%%%%%%%%%%%%%%%%%%%%%%%%%%%%%
%\section{Acknowledgements}
\textbf{Acknowledgements}
%%%%%%%%%%%%%%%%%%%%%%%%%%%%%%%%%%%%%%%%%%%%%%%%%%%%%%%%%%%%%%%%%%%%
%%%%%%%%%%%%%%%%%%%%%%%%%%%%%%%%%%%%%%%%%%%%%%%%%%%%%%%%%%%%%%%%%%%%

%This work is partly supported by the Brazilian agencies CNPq (L.M.A.: Grant Numbers 309950/2020-1, 400215/2022-5, 200567/2022-5), FAPESP (K.P.K.: Grant Number 2022/08347-9; A. M. T.: Grant number 2023/01182-7); FAPESB (L.M.A.: Grant Number INT0007/2016); and CNPq/FAPERJ under the Project INCT-F\'{\i}sica Nuclear e Aplicações (Contract No. 464898/2014-5).

This work is partly supported by the Brazilian agencies CNPq, FAPESP, FAPESB and CNPq/FAPERJ under the Project INCT-F\'{\i}sica Nuclear e Aplicações.

\bibliographystyle{apsrev4-1}
\bibliography{D1refs}

\end{document}